\documentclass[12pt]{article}
\input epsf.sty
\usepackage{mathrsfs}
\usepackage{amsmath}
\usepackage{mathrsfs}
\usepackage{amssymb}
\usepackage{color}
\usepackage{psfrag}

\newcommand{\bea}{\begin{eqnarray}}
\newcommand{\eea}{\end{eqnarray}}
\newcommand{\be}{\begin{equation}}
\newcommand{\ee}{\end{equation}}
\newcommand{\vs}[1]{\vspace{#1 mm}}

\renewcommand{\b}{\beta}

\newcommand{\dsl}{\pa \kern-0.5em /}

\newcommand{\pa}{\partial}

\newcommand{\nn}{\nonumber\\}

\newcommand{\eqn}[1]{(\ref{#1})}

\begin{document}
\topmargin 0pt
\oddsidemargin 0mm

\begin{flushright}

USTC-ICTS-10-12\\


\end{flushright}

\vspace{2mm}

\begin{center}

{\Large \bf Phase structure of black branes in \\
grand canonical ensemble}

\vs{10}

{\large J. X. Lu$^{a, b}$\footnote{E-mail: jxlu@ustc.edu.cn},
Shibaji Roy$^c$\footnote{E-mail: shibaji.roy@saha.ac.in} and
Zhiguang Xiao$^a$\footnote{E-mail: xiaozg@ustc.edu.cn}}

 \vspace{4mm}

{\em

 $^a$ Interdisciplinary Center for Theoretical Study\\
 University of Science and Technology of China, Hefei, Anhui
 230026, China\\

\vs{4} $^b$ Kavli Institute for Theoretical Physics China, CAS,
Beijing 100190, China \vs{4}

 $^c$ Saha Institute of Nuclear Physics,
 1/AF Bidhannagar, Calcutta-700 064, India\\}

\end{center}

\vs{10}

\begin{abstract}

This is a companion paper of our previous work [1] where we studied
the thermodynamics and phase structure of asymptotically flat black
$p$-branes in a cavity in arbitrary dimensions $D$ in a canonical
ensemble. In this work we study the thermodynamics and phase
structure of the same in a grand canonical ensemble. Since the
boundary data in two cases are different (for the grand canonical
ensemble boundary potential is fixed instead of the charge as in
canonical ensemble) the stability analysis and the phase structure
in the two cases are quite different. In particular, we find that
there exists an analog of one-variable analysis as in canonical
ensemble, which gives the same stability condition as the rather
complicated known (but generalized from black holes to the present case)
two-variable analysis. When certain condition for the
fixed potential is satisfied, the phase structure of charged black
$p$-branes is in some sense similar to that of the zero charge black
$p$-branes in canonical ensemble up to a certain temperature. The
new feature in the present case is that above this temperature,
unlike the zero-charge case, the stable brane phase no longer exists
and `hot flat space' is the stable phase here. In the
grand canonical ensemble there is an analog of Hawking-Page
transition, even for the charged black $p$-brane, as opposed to the
canonical ensemble. Our study applies to non-dilatonic as well as
dilatonic black $p$-branes in $D$ space-time dimensions.

\end{abstract}

\newpage

\section{Introduction}

Black holes in asymptotically AdS space have attracted a lot of
attention in recent years. The reason is partly due to the AdS/CFT
correspondence proposed by Maldacena \cite{Maldacena:1997re} and its
consequences \cite{Witten:1998qj,Gubser:1998bc,Aharony:1999ti}.
The large black holes in AdS space are thermodynamically
stable\footnote{This is not the case for small black holes, for
example, see \cite{Horowitz:1999jd}.} \cite{Hawking:1982dh} and so,
the equilibrium states and the phase structure of such space-times
can be easily studied. By AdS/CFT correspondence this in turn will
provide us with the information about the similar states and phase
structure on the CFT or gauge theory side
\cite{Chamblin:1999tk,Chamblin:1999hg}. Black hole space-time is
naturally associated with a Hawking temperature
\cite{Hawking:1974sw}, so the gauge theory will also be at a finite
temperature \cite{Witten:1998zw}. In particular, AdS black holes are
well-known to undergo a Hawking-Page transition
\cite{Hawking:1982dh} and by AdS/CFT, this corresponds to the
confinement-deconfinement phase transition \cite{Witten:1998zw} in
SU($N$) gauge theory at large $N$.

However, it is well-known that the phase structure just
mentioned is not unique
to the AdS black holes, but similar structure also arises
in suitably
stabilized flat as well as dS black holes
\cite{Carlip:2003ne,Lundgren:2006kt}.
Higher dimensional theories like
string or M-theory are known to admit higher dimensional black
objects in the form of
black $p$-branes \cite{Horowitz:1991cd,Duff:1993ye,Duff:1996hp}
which are asymptotically flat and so, it would be interesting
to see what kind of equilibria and phase structure they give rise
to. With this motivation we obtained the equilibrium states and the phase
structure of the black $p$-branes in the canonical ensemble in the previous
work [1]. In this paper we study the same for the black $p$-branes in
grand canonical ensemble. Since the boundary data in these two ensembles are
quite different, their phase structures are also expected to be different and
it is of interest to see how they differ with the change of the boundary data.

We will consider the black $p$-branes (in $D$ dimensions) which are
the solutions of $D$-dimensional gravity coupled to a dilaton (a
scalar field) and a $(p+1)$-form gauge field \cite{Duff:1993ye}.
These are asymptotically flat and so are thermodynamically
unstable\footnote{A detailed analysis of this kind of
instability for neutral black holes/branes/rings is given in
\cite{Dias:2010eu, Astefanesei:2010bm}.} as an isolated black
$p$-brane would radiate energy in the form of Hawking radiation
\cite{Hawking:1974sw}.  In order to restore thermodynamic stability
so that equilibrium thermodynamics and phase structure can be
studied we must consider ensembles that include not only the branes
but also their environment. As self-gravitating systems are
spatially inhomogeneous, any specification of such ensembles
requires not just thermodynamic quantities of interest but the place
at which they take the specified values. In other words, we place
the black brane in a cavity {\it a la} York \cite{York:1986it}
 (see also \cite{Brown:1994sn,
Parentani:1994wr,Peca:1998dv,Gregory:2001bd,Zaslavskii:2003is}) and
its extension in the charged case. The wall of the cavity is fixed
at a radial distance $\rho_B$, and we will keep the temperature and
the gauge potential at the wall of the cavity (at $\rho_B$) fixed.
This will define a grand canonical ensemble (Note that for the
canonical ensemble, the charge enclosed in the cavity is fixed
instead.). So, for the grand canonical ensemble charge can vary. We
will study the phase structure of black $p$-branes in this ensemble.
Charged black holes in the grand canonical ensemble have been
studied in \cite{Whiting:1988qr,Braden:1990hw}.

We will employ the Euclidean action formalism
\cite{Gibbons:1976ue,York:1986it,Brown:1994su} to the dilatonic
black $p$-brane geometry in $D$ space-time dimensions. We first
write the Euclidean action containing the gravitational part
including a Gibbons-Hawking boundary term \cite{Gibbons:1976ue}, the
dilatonic part and the form-field part. By using the equation of
motion(i.e., on-shell), we can express the usual Einstein-Hilbert
gravitational action plus the dilatonic part in terms of the
form-field, therefore simplify the action and then evaluate it for
the given black $p$-brane geometry placed in a cavity with the
potential and the temperature fixed on its wall.  To the leading
order this action is related to the grand potential or Gibbs free
energy of the system and is an essential entity for the stability
analysis. It can explicitly be expressed in terms of parameters of
the branes and the cavity. The stationary point of this action with
respect to the relevant variables such as the horizon size will
determine the relevant thermal states with the temperature and
potential fixed at the wall of the cavity. The second derivative or
derivatives of the action at the stationary point can be used to
give us the information about the stability of the black brane at
that point. We find the condition which determines at least the
local stability of the black brane phase. In the case when there is
a stable black brane phase, we find that there exists a minimum
temperature below which there is only `hot flat space' phase and no
black brane phase. But above this temperature there exist two black
brane phases with two different radii. The smaller one is unstable
which corresponds to the maximum of the free energy and the
larger one is locally stable and corresponds to the local minimum of
the free energy. This local minimum becomes a global one only after
the temperature rises above a certain value. Below this value and
above minimum temperature, the locally stable black brane eventually
makes a transition to the `hot flat space' by a  phase transition.
But above this value, the large black brane is globally stable. Upto
this point the picture is very similar to the chargeless black brane
case in the canonical ensemble [1]. But now for the grand canonical
ensemble as the temperature rises more, the globally stable black
brane phase disappears after a certain value and at this point there
is only one unstable black brane phase and the stable phase here
would be the `hot flat space'. Also note that unlike in the case of
canonical ensemble\footnote{For the case of canonical ensemble with
fixed non-zero charge,
 the phase structure of the asymptotically flat black $p$-branes
\cite{Lu:2010xt}, when $\tilde d > 2$, with $\tilde d = D - 3 - p$,
resembles the one in \cite{Chamblin:1999tk, Chamblin:1999hg} for
charged AdS black holes as discussed in detail in \cite{Lu:2010xt}.
For examples, in both cases, there exist a critical charge and when
the charge carried by the branes is greater than this charge, there
always exists a thermodynamically stable state for each given
temperature. When the charge is less than the critical value, there
exists a line of first order phase transition of van der
Waals-Maxwell liquid-gas type, ending at a second order phase
transition point when the charge reaches the critical value. The
critical exponent for the specific heat is universal and has the
value $- 2/3$. However, for $\tilde d = 1$ the phase structure is
similar to the chargeless case except that there does not appear to
exist an analog of Hawking-Page transition due to the presence of
charge and the horizon size has a lower bound set by the charge
(rather than zero as in the chargeless case). The $\tilde d = 2$
case is a borderline between $\tilde d = 1$ case and $\tilde d > 2$
cases. Similar to the cases of $\tilde d > 2$, it has a critical
charge and three subcases depending on whether its charge is greater
than, equal to or less than the critical charge. However, there
exists neither a critical behavior nor a Van der Waals-Maxwell type
liquid-gas phase transition. The stable phase structure looks more
like the $\tilde d = 1$ case. For other details we refer the reader
to \cite{Lu:2010xt}. } \cite{Chamblin:1999tk, Chamblin:1999hg,
Carlip:2003ne, Lundgren:2006kt, Lu:2010xt}, here we do not have a
situation where we have two locally stable black brane phases and so
there is no such phase transition similar to the van der
Waals-Maxwell liquid-gas phase transition or even a transition from
one black brane phase to the other. We would like to remark that in
this grand canonical ensemble, unlike in the case of AdS black holes
\cite{Chamblin:1999tk, Chamblin:1999hg}, we do not have a critical
value for the potential in the sense defined there since the
existence of this critical value in AdS case is largely due to the
presence of the cosmological constant ( as an additional parameter).
One can check easily that if we set the cosmological constant vanish
(i.e., setting the parameter $l \to \infty$ there), the extremal
black holes becomes  supersymmetric BPS ones, which is in analog of
the present discussion, and the potential will not have the critical
value anymore but has only an upper bound. The existence of the
upper bound of potential for the present case is explained below and
in section 3.

We now discuss the case of extremal black $p$-branes.  Mathematically,
the extremal brane configuration can be obtained simply by taking the
`mass-equal-charge' limit of the corresponding non-extremal one. However,
this does not mean that the extremal branes and non-extremal branes can
actually be related to each other by
a physical process which can also be understood from the semi-classical
discussion given by Gibbons and Kallosh \cite{Gibbons:1994ff},
Hawking {\it et al} \cite{Hawking:1994ii, Hawking:1995fd} and  Teitelboim
\cite{Teitelboim:1994az}, for the case of black holes including the
dilatonic ones (the latter resembles the present discussion). So, for example,
the entropy changes discontinuously from the non-extremal case to the extremal one.
 The non-extremal case has a non-vanishing entropy,
given always by one quarter of the event horizon area while the extremal one
always has vanishing entropy\footnote{The discrepancy
in the entropy between the semi-classical
calculation and the microstate counting in string theory in the
AdS/CFT context has been addressed recently in
\cite{Carroll:2009maa}. It was pointed out there that the
non-vanishing of the entropy is due to a different reason and the
semi-classical result of vanishing entropy for the extremal black hole
still holds.}. This very fact, just like the case of black holes,
implies that the non-extremal branes and extremal branes  are
qualitatively different objects and a non-extremal black brane cannot
turn into an extremal brane. Actually, the nearer the mass of the non-extremal
brane gets to the charge the lower the temperature of the brane and so
the lower the rate of radiation of mass. Thus the mass will never be exactly
equal to the charge.
This, in turn, implies, as it should, that there does not exist any real
physical process which can bring down the temperature
of a thermal system to absolute zero since otherwise it would be against the
third law of thermodynamics that the zero temperature can never be
reached in reality. Moreover, also like the case of black holes, an
extremal brane cannot become a non-extremal brane even when matter or
radiation are thrown into
the extremal one, which seems at first sight contrary to the common
sense \cite{Hawking:1994ii}. This is due to the fact that one can identify
extremal branes with any
period and therefore they can be in equilibrium with thermal radiation
at any temperature. Thus they must be able to radiate at any rate, unlike
non-extremal branes, which can radiate only at the rate corresponding to their
temperature. So extremal objects always radiate in such a way as to keep themselves
extreme when matter or radiation is sent into them. This discussion applies to
the asymptotically flat branes in canonical ensemble for which the charge
inside the cavity is fixed.

In this paper, we are considering the asymptotically flat branes in
grand canonical ensemble for which the potential at the wall of
the cavity is fixed instead.
Before we discuss the thermodynamical stability, we first need to address
the question: given the boundary data in this ensemble mentioned earlier,
what are the possible stable configurations allowed in the cavity which
can be used to discuss this stability? Unlike in the canonical ensemble,
the charge inside
the cavity in the grand canonical ensemble is not fixed and can fluctuate
or exchange with the environment. So one expects first that the `hot flat
space' can be a stable phase under certain conditions. The second candidate
is the non-extremal branes, charged or uncharged, and the third possible
one is the extremal branes. However, the probability for the existence of
non-extremal branes is much higher than that of the extremal branes for the
following reasons. In the case of black holes, the pair creation rate of
the non-extremal black holes is enhanced by a factor of $e^S$ over that of
the extremal branes where $S$ is the entropy of the non-extremal black holes
\cite{Hawking:1994ii, Gibbons:1994ff}. We expect this to apply to the case
of branes as well. Further, unlike  a non-extremal brane, an extremal brane
with a large charge can fission (split) into extremal branes with smaller
charges without violating the second law of thermodynamics since this process
is suppressed by a factor of $e^{\Delta S}$ with $\Delta S$ the change of
total entropy of the system of extremal branes and for extremal branes
$\Delta S = 0$. Note that an extremal brane with a macroscopic charge in the
present context (for the gravity description to remain valid) is actually a
BPS brane, preserving one half of the underlying spacetime supersymmetries
and there is no binding energy among BPS branes with smaller charges,
even the individual BPS branes with unit charge (which constitutes the extremal
branes with macroscopic charge). So a large extremal branes can easily
split into smaller extremal ones. This provides another means to understand
the above fission process. In other words, the extremal branes are unstable
in the first place in this ensemble and should be excluded from consideration
in our following thermodynamical stability analysis\footnote{So the natural
background in the grand canonical ensemble in the present discussion is
the `hot flat space' while from the previous discussion the natural background
in the canonical ensemble is the extremal branes.}.

This paper is organized as follows. In section 2, we discuss the dilatonic
black $p$-brane solution and evaluate the action in Euclidean
signature. Then we discuss the stability of various equilibrium states from
this action in section 3. The phase structure of the black $p$-brane is
discussed in section 4. Then we conclude in section 5. A more
general two-variable stability analysis (as opposed to one-variable
stability analysis performed in section 3) is presented in
the Appendix.

\section{Black $p$-brane solution and the action}

The black $p$-brane solution was originally constructed
\cite{Horowitz:1991cd} as a solution to the ten dimensional
supergravity containing a metric, a dilaton and a $(p+1)$-form gauge
field and was generalized to arbitrary dimensions in
\cite{Duff:1993ye}. These solutions are given in Lorentzian
signature, but for the purpose of studying thermodynamics, we write
the black $p$-brane solution in Euclidean signature as (see for
example \cite{Duff:1994an}), \bea
 \label{blackbrane-phi}
d s^2 &=& \Delta_+\Delta_-^{- \frac{d}{D - 2}} d t^2 +
    \Delta_-^{\frac{\tilde d}{D - 2}} \sum_{i=1}^{d-1}(d x^i)^2 +
\Delta_+^{-1}
    \Delta_-^{\frac{a^2}{2\tilde d} - 1} d\rho^2 + \rho^2
\Delta_-^{\frac{a^2}{2\tilde d}} d \Omega_{\tilde d +
    1}^2,\nn
    A_{[p+1]} &=& - i e^{a\phi_0/2}
\left[\left(\frac{r_-}{r_+}\right)^{\tilde
    d/2} - \left(\frac{r_- r_+}{\rho^2}\right)^{\tilde
    d/2}\right] dt \wedge dx^1\wedge \ldots \wedge dx^p,\nn
    F_{[p+2]} &\equiv&  dA_{[p+1]}
    = - i e^{a\phi_0/2} \tilde d \,
\frac{(r_- r_+)^{\tilde d / 2}}{\rho^{\tilde d  +
    1}} d\rho \wedge dt \wedge dx^1 \wedge \ldots \wedge dx^p,
\nn e^{2(\phi-\phi_0)}&=& \Delta_-^a, \eea Here we have defined
\be\label{deltapm} \Delta_{\pm} = 1 -
\left(\frac{r_{\pm}}{\rho}\right)^{\tilde d} \ee where, $r_{\pm}$
are the two parameters characterizing the solution and are related
to the mass and the charge of the black brane. In the metric
\eqn{blackbrane-phi} the Euclidean time is periodic and so, the
metric has an isometry S$^1$ $\times$ SO($d-1$) $\times$ SO($\tilde
d +2$) indicating that it represents a $(d-1) \equiv p$-brane in
Euclidean signature. The total space-time dimension is $D=d+\tilde d
+2$, where the space transverse to the $p$-brane has the
dimensionality $\tilde d +2$. $\phi$ is the dilaton and $\phi_0$ is
its asymptotic value and related to the string coupling as $g_s =
e^{\phi_0}$. $a$ is the dilaton coupling and is given for the
supergravity theory with maximal supersymmetry by, \be\label{a} a^2
= 4 - \frac{2d\tilde d}{D-2}. \ee It is clear from the Lorentzian
form of the above metric that when $r_- =0$, and $a=0$, it reduces
to the $D$-dimensional Schwarzschild solution which has an event
horizon at $\rho=r_+$, whereas, at $\rho = r_-$, there is a
curvature singularity. So, the metric in \eqn{blackbrane-phi}
represents a black $p$-brane only for $r_+>r_-$, with $r_+ = r_-$,
being its extremal limit \cite{Horowitz:1991cd}. A $p$-brane
naturally couples to the $(p+1)$-form gauge field whose form and its
field strength are given in \eqn{blackbrane-phi}. Note that we have
defined the gauge potential with a constant shift, following
\cite{Braden:1990hw}, in such a way that it vanishes on the horizon
so that it is well-defined on the local inertial frame. The black
$p$-brane will be placed in a cavity with its wall at $\rho =
\rho_B$. It is clear from the metric in \eqn{blackbrane-phi} that
the physical radius of the cavity is \be\label{rhophysical}
\bar\rho_B = \Delta_-^{\frac{a^2}{4 \tilde d}} \rho_B \ee while
$\rho_B$ is merely the coordinate radius. So, it is $\bar\rho_B$
which we should fix in the following discussion and not $\rho_B$.
Note that when the black brane is non-dilatonic $a=0$ and in that
case $\rho_B = \bar \rho_B$. Also we fix the dilaton \footnote{This
enables us to obtain its correct equation of motion from the
corresponding action in the presence of a boundary. } at
$\bar\rho_B$, which indicates that the asymptotic value of the
dilaton is not fixed for the present consideration and this is
crucial for our discussion in expressing relevant quantities in
`barred' parameters. By this argument we also have \be\label{rpm}
\bar r_{\pm} = \Delta_-^{\frac{a^2}{4\tilde d}} r_{\pm} \ee and
$\bar r_{\pm}$ are the proper parameters which we should use in the
present context. In terms of the `barred' parameters $\Delta_{\pm}$
remain the same as before, \be\label{deltanpm} \Delta_{\pm} = 1 -
\frac{r_{\pm}^{\tilde d}}{\rho_B^{\tilde d}} = 1 - \frac{\bar
r_{\pm}^{\tilde d}}{\bar \rho_B^{\tilde d}}. \ee Since the Euclidean
time coordinate in \eqn{blackbrane-phi} is periodic so, for the
metric to be well-defined without a conical singularity at $\rho =
r_+$, the Euclidean time must have a periodicity,
\be\label{beta-phi} \b^{\ast} = \frac{4\pi r_+}{\tilde d}\left(1 -
\frac{r_-^{\tilde
      d}}{r_+^{\tilde d}}\right)^{\frac{1}{\tilde d} - \frac{1}{2}},
\ee which is the inverse temperature at $\rho = \infty$. The local
$\b (\bar\rho_B) $ is given\footnote{The corresponding local
temperature is related to the temperature at spatial infinity by the
so-called Tolman relation \cite{Tolman:1930zz}.} as
\be\label{localbeta} \b = \b (\bar\rho_B) = \Delta_+^{\frac{1}{2}}
\Delta_-^{-\frac{d}{2(d+\tilde d)}} \b^{\ast} \ee which is the
inverse of local temperature at $\bar\rho_B$ when in thermal
equilibrium with the environment (the wall of the cavity) at the
inverse temperature $\b$. Note that in terms of the `barred'
parameters the inverse of local temperature $\b (\bar \rho_B)$ can be
expressed from \eqn{localbeta} and \eqn{rhophysical} as,
\be\label{betainbar} \b (\bar \rho_B)= \frac{4\pi r_+}{\tilde d}
\Delta_+^{\frac{1}{2}} \Delta_-^{-\frac{d}{2(d+\tilde d)}}
\left(1-\frac{r_-^{\tilde d}}{r_+^{\tilde
      d}}\right)^{\frac{1}{\tilde d} - \frac{1}{2}} =
\frac{4\pi \bar r_+}{\tilde d} \Delta_+^{\frac{1}{2}}
\Delta_-^{-\frac{1}{\tilde d}} \left(1-\frac{\bar r_-^{\tilde
d}}{\bar r_+^{\tilde
      d}}\right)^{\frac{1}{\tilde d} - \frac{1}{2}},
\ee where in the second equality $\Delta_\pm$ are also expressed
in terms of `barred' parameters.
 Also the charge is defined as,
\bea\label{charge-phi} Q_d &=& \frac{i}{\sqrt{2}\kappa}\int
e^{-a(d)\phi} \ast F_{[p+2]} = \frac{\Omega_{\tilde d
+1}}{\sqrt{2}\kappa}e^{-a\phi_0/2}\tilde d (r_+ r_-)^{\tilde d/2}\nn
&=& \frac{\Omega_{\tilde d +1} \tilde d}{\sqrt{2}\kappa} e^{-a\bar
\phi/2} (\bar r_+ \bar r_-)^{\tilde d/2}. \eea In \eqn{charge-phi},
$\kappa$ is a constant with $1/ (2 \kappa^2)$ appearing in front of
the Hilbert-Einstein action in canonical frame but containing no
string coupling $g_s$, $\ast F_{[p+2]}$ denotes the Hodge dual of
the $(p+2)$-form field given in \eqn{blackbrane-phi}. Also,
$\Omega_n$ denotes the volume of a unit $n$-sphere. In the last line
of \eqn{charge-phi} we have expressed the asymptotic value of the
dilaton in terms of the fixed dilaton via the boundary condition
$\phi(\bar \rho_B) = \bar \phi$ at the wall of the cavity from the
relation \eqn{blackbrane-phi} and then expressed $r_{\pm}$ by $\bar
r_{\pm}$ from \eqn{rpm}.

In the grand canonical ensemble, the fixed quantities are the
physical radius of the wall of the
 cavity $\bar \rho_B$, the temperature $T$, the brane volume $ V_p$
(The local volume at $\bar\rho_B$, $V_p (\bar\rho_B) =
\Delta_-^{\frac{\tilde d
    (d-1)}{2(D-2)}} V_p^{\ast}$, with $V_p^{\ast} = \int d^p x$,
 is set equal to the fixed value $ V_p$) and the
potential $\bar \Phi$, all at the wall of the cavity. They are
independent parameters and the corresponding local quantities for
the branes have to be equal to these pre-selected fixed values on-shell
at the wall of the cavity, respectively. The potential in the local
inertial frame at the wall of the cavity can be obtained from
$A_{[p+1]}$ and the metric in \eqn{blackbrane-phi} as,
\bea\label{potential} A_{[p+1]} &=& -i e^{a\bar \phi/2}
\left(\Delta_- \Delta_+\right)^{-\frac{1}{2}}
\left(\frac{r_-}{r_+}\right)^{\frac{\tilde d}{2}} \left(1 -
\frac{r_+^{\tilde d}}{\rho_B^{\tilde d}}\right) d\bar t \wedge d\bar
x^1 \ldots \wedge d\bar x^p\nn &\equiv & -i \sqrt{2} \kappa \Phi
d\bar t \wedge d\bar x^1 \ldots d\bar x^p \eea where $(\bar t, \,\,
\bar x^1,\,\, \ldots, \bar x^p)$ are the coordinates in the local
inertial frame and is related to the original coordinates as $\bar t
= \Delta_+^{\frac{1}{2}}\Delta_-^{-\frac{d}{2(D-2)}} t$ and $\bar
x^i = \Delta_-^{\frac{\tilde d}{2(D-2)}} x^i$ for $i=1,\,\,2,\ldots,
p$ as can be seen from the metric in \eqn{blackbrane-phi}. So,
$\Phi$ is the potential conjugate to the charge and is set equal to
the fixed value $\bar \Phi$ at the wall. We will now evaluate the
action with these boundary data.

The Euclidean action for the dilatonic black branes in the canonical
ensemble has already been evaluated in the Appendix of our previous
work [1]. We will use that result to evaluate the action for the
grand canonical ensemble. The relevant action for the gravity
coupled to the dilaton and a $(p+1)$-form gauge field in the
canonical ensemble is, \be\label{action} I^C_E = I^C_E(g) +
I^C_E(\phi) + I^C_E(F) \ee where, $I^C_E(g)$ is the gravitational
part of the action which has a Hilbert-Einstein term and a
Gibbons-Hawking boundary term \cite{Gibbons:1976ue}, $I^C_E(\phi)$
is the dilatonic part and $I^C_E(F)$ is the form-field part. The
first two terms remain exactly the same in the grand canonical
ensemble, however, the form-field part which in the canonical
ensemble has the form, \bea\label{formfieldaction} I^C_E (F) &=&
\frac{1}{2 \kappa^2} \frac{1}{2 (d + 1)!} \int_M
   d^D x \sqrt{g}\,e^{-a(d)\phi}\, F^2_{d + 1}
\nonumber\\
&&- \frac{1}{2 \kappa^2}
   \frac{1}{d!} \int_{\partial M} d^{D - 1} x \sqrt{\gamma}\, n_\mu
   \,e^{-a(d)\phi}\, F^{ \mu \mu_1 \mu_2 \cdots \mu_d} A_{\mu_1 \mu_2 \cdots
   \mu_d},
\eea will differ in the grand canonical ensemble since the potential
at the wall of the cavity is fixed and so, the last term in
\eqn{formfieldaction} will be absent in the grand canonical
ensemble. Therefore, the action for the grand canonical ensemble can
be obtained from that of the canonical ensemble by the relation,
\be\label{ctogcaction} I^{GC}_E = I^C_E +
\frac{1}{2\kappa^2}\frac{1}{d!} \int_{\partial M} d^{D - 1} x
\sqrt{\gamma}\, n_\mu \,e^{-a(d)\phi}\, F^{ \mu \mu_1 \mu_2 \cdots
\mu_d} A_{\mu_1 \mu_2 \cdots\mu_d}. \ee Here $\partial M$ denotes
the boundary of the space-time $M$ and $n_\mu$ is a space-like
vector normal to the boundary. $\gamma$ is the determinant of the
boundary metric. Note that the spacetime $M$ here is the bulk region
defined by $\bar\rho_B \ge \bar \rho \ge \bar r_+$ while the
boundary $\partial M$ is defined as the surface at $\bar \rho =
\bar\rho_B$ plus the one at $\bar\rho = \bar r_+$. The surface
integration in \eqn{ctogcaction} at $\bar\rho = \bar r_+$ vanishes
due to the vanishing form-potential there.  Evaluating the last term
in \eqn{ctogcaction} for the black $p$-brane configuration given in
\eqn{blackbrane-phi} we get, \be\label{ctogcaction1} I^{GC}_E =
I^C_E - \frac{\b^{\ast} V_p^{\ast}}{2\kappa^2} \Omega_{\tilde d+1}
\tilde d r_-^{\tilde d} \Delta_+ = I^C_E - \b V_p Q_d \bar \Phi, \ee
where we have used the charge expression and the potential from
\eqn{charge-phi} and \eqn{potential} as, \bea\label{chargephi} Q_d
&=& \frac{\Omega_{\tilde d +1} \tilde d}{\sqrt{2}\kappa} e^{-a\bar
\phi/2} (\bar r_+ \bar r_-)^{\frac{\tilde d}{2}}, \nn \bar \Phi &=&
\Phi (\bar \rho_B) = \frac{1}{\sqrt{2}\kappa} e^{a\bar \phi/2}
\left(\frac{\bar r_-}{\bar
    r_+}\right)^{\frac{\tilde d}{2}}
\left(\frac{\Delta_+}{\Delta_-}\right)^{\frac{1}{2}}. \eea Now using
the explicit form\footnote{In obtaining this, we have made use of
the on-shell equation of motion  \be R - \frac{1}{2} (\partial
\phi)^2 = \frac{\tilde d - d}{2 (D - 2) (d + 1)!} e^{- \alpha (d)
\phi} F^2,\ee to express $I_E (g) + I_E (\phi)$ in terms of the
form-field $F$ to simplify the computation.} of the Euclidean action
for the canonical ensemble given in [1] we write \eqn{ctogcaction1}
as\footnote{In obtaining the explicit action, we have used the
background subtraction approach. The natural background here is
taken as the `hot flat space' with a constant dilaton $\bar\phi$, a
fixed potential $\bar \Phi$ and other boundary data. The equations
of motion are satisfied trivially inside the cavity for the `hot
flat space' with the given boundary data.} \bea\label{ctocgaction2}
I_E(\b,\bar\Phi,\bar \rho_B, V_p;Q_d,\bar r_+) &=& \beta  \Omega \nn
&=&\beta E (\bar \rho_B, V_p; Q_d, \bar r_+) - S (\bar\rho_B, V_p;
Q_d, \bar r_+) - \b V_p Q_d \bar\Phi\nn &=& -\frac{\beta V_p
\Omega_{\tilde d + 1}}{2 \kappa^2} \bar \rho_B^{\tilde d}
\left[(\tilde d + 2) \left(\frac{\Delta_+}{\Delta_-}\right)^{1/2} +
\tilde d (\Delta_+ \Delta_-)^{1/2} - 2 (\tilde d + 1)\right]\nn & &
-\frac{4\pi V_p \Omega_{\tilde d +1}}{2\kappa^2} \bar r_+^{\tilde d
+1} \Delta_-^{-\frac{1}{2} - \frac{1}{\tilde d}} \left(1-\frac{\bar
r_-^{\tilde d}}{\bar r_+^{\tilde d}}\right)^{\frac{1}{2} +
  \frac{1}{\tilde d}} - \b V_p Q_d \bar\Phi,
\eea where $\Omega = E - T S - Q \bar\Phi$ is the Gibbs free energy.
In the above, we have the total charge  enclosed in the cavity as $Q
= V_p Q_d$, the entropy \bea S &=& \frac{4\pi V_p \Omega_{\tilde d
+1}}{2\kappa^2} \bar r_+^{\tilde d +1} \Delta_-^{-\frac{1}{2} -
\frac{1}{\tilde d}} \left(1-\frac{\bar r_-^{\tilde d}}{\bar
r_+^{\tilde d}}\right)^{\frac{1}{2} +
  \frac{1}{\tilde d}},\nn
  &=& \frac{4\pi V_p^{\ast}\Omega_{\tilde d+1}}{2\kappa^2} r_+^{\tilde d +1}
\left(1-\frac{r_-^{\tilde d}}{r_+^{\tilde d}}\right)^{\frac{1}{2} +
  \frac{1}{\tilde d}},\eea where $V_p^* = \int d^p x$ as defined before
and is independent of
   the location of the cavity as it should be, and the energy for the cavity as
   \be E = -
\frac{ V_p \Omega_{\tilde d + 1}}{2 \kappa^2} \bar \rho_B^{\tilde d}
\left[(\tilde d + 2) \left(\frac{\Delta_+}{\Delta_-}\right)^{1/2} +
\tilde d (\Delta_+ \Delta_-)^{1/2} - 2 (\tilde d + 1)\right],\ee
which gives the ADM mass when $ \rho_B \to \infty$. Since
this is the Euclidean action for the black $p$-brane in the grand
canonical ensemble we will be working with, for brevity, we have
removed the superscript `$GC$' from $I_E$ in writing
\eqn{ctocgaction2}. Note that we have expressed everything on the
r.h.s. in terms of the `barred' parameters showing that the
formalism works for both the non-dilatonic as well as dilatonic
branes.  In \eqn{ctocgaction2} $\b$, $\bar \rho_B$, $\bar\Phi$ are
the inverse of temperature, the physical radius and the potential of
the cavity, therefore are all fixed, and $Q_d$, $\bar r_+$ are
variables. Note that $\bar r_-$ is not independent and can be
expressed in terms of $Q_d$ and $\bar r_+$ from \eqn{chargephi} as,
\be\label{rminus} \bar r_-^{\tilde d} = \left(\frac{\sqrt{2}\kappa
Q_d}{\Omega_{\tilde d+1}
    \tilde d} e^{a\bar \phi/2}\right)^2 \frac{1}{\bar r_+^{\tilde
    d}}.
\ee Substituting \eqn{rminus} in \eqn{ctocgaction2} we can express
the action in terms of two variables $Q_d$ and $\bar r_+$. Then
varying the action with respect to $Q_d$ and putting that to zero,
i.e., at the stationary point we obtain (after some simplification),
\bea\label{didq} && \frac{\partial I_E}{\partial Q_d} = 0\nn &&
\Rightarrow \quad \bar\Phi = \Phi (\bar\rho_B) = \frac{Q_d e^{a\bar
\phi}}{\tilde d \Omega_{\tilde
    d +1} \bar r_+^{\tilde
    d}}\left(\frac{\Delta_+}{\Delta_-}\right)^{\frac{1}{2}} \left[1 +
    \frac{\tilde d +2}{\tilde d} \Delta_-^{-1}\left(\frac{4\pi \bar r_+
        \Delta_+^{\frac{1}{2}}\Delta_-^{-\frac{1}{\tilde d}}}{\b \tilde d
\left(1 - \frac{\bar r _-^{\tilde d}}{\bar r_+^{\tilde d}}\right)^{\frac{1}{2}
  -\frac{1}{\tilde d}}} - 1 \right)\right].
\eea Similarly, varying the action with respect to the other
variable $\bar r_+$ and setting that to zero, i.e., at the
stationary point we obtain, \be\label{didrplus} \frac{\partial
I_E}{\partial \bar r_+} = 0 \quad \Rightarrow \quad \b = \b
(\bar\rho_B) = \frac{4\pi \bar r_+}{\tilde d} \Delta_+^{\frac{1}{2}}
\Delta_-^{-\frac{1}{\tilde d}}\left(1 - \frac{\bar r_-^{\tilde
d}}{\bar
    r_+^{\tilde d}}\right)^{\frac{1}{\tilde d} - \frac{1}{2}}.
\ee
This is the correct form of $\b$ we had given earlier in
\eqn{betainbar}. Using this \eqn{didrplus} in \eqn{didq} and substituting the
form of $Q_d$ given in the first equation of \eqn{chargephi} we recover the
correct form of the potential given in the second equation of \eqn{chargephi}.
This is a verification that the action $I_E$ given in \eqn{ctocgaction2} is
indeed correct. We will use this form of the action in the next section to
study the stability of the equilibrium states of the black $p$-branes in the
grand canonical ensemble.

Before we close this section, we would like to discuss briefly the
validity of the use of the effective action in describing the phase
structure of black $p$-branes throughout the parameter space in the
present ensemble.  Since we are considering a situation where the black
$p$-branes are placed in a cavity, this implies that the size of the
cavity $\bar
\rho_B$ must be larger than the horizon size $\bar r_+$ of the
branes. In obtaining the explicit Euclidean action
\eqn{ctocgaction2}, we have performed the integration of the bulk
action in the range of $\bar \rho_B
> \bar \rho > \bar r_+$. To have a valid description in terms of
the effective action,
we need to keep the curvature of black brane spacetime uniformly
weak throughout this range and if the fundamental theory is string
theory, we need in addition to keep the effective string coupling
uniformly weak. One can check that the requirement(s) can be
satisfied with the expected condition \be \label{weakc} \bar r_+ =
r_+ \Delta_-^{a^2/4\tilde d} \gg l,\ee where $l$ is the fundamental
length scale of the underlying theory, for example,
 it is the Planck scale $l_p$ in eleven dimensions or the string
scale $l_s$ in ten dimensions.

There appears no further subtlety
in the grand canonical ensemble considered in this paper since the extremal
branes are excluded from consideration as discussed in the Introduction.
The relevant configurations here are the `hot flat space' and the
non-extremal branes. Note also that the location of the cavity wall
along the radial direction satisfies
$\bar\rho_B > \bar\rho > \bar r_+ > \bar r_-$. However, there
appears to have a potential issue regarding the fixed dilaton
value at the wall of the cavity in
the canonical ensemble, considered in our previous work \cite{Lu:2010xt},
for certain extremal branes such as D3 branes for which the so-called
attractor mechanism is at work\footnote{
We thank the anonymous referee for raising this point to us.}. Once this
mechanism is at work, the dilaton value at the horizon is completely
determined by the relevant charge(s), independent of its asymptotic value.
The extremal $p$-brane configuration from the supergravities with maximal
SUSY in spacetime dimension $D \ge 4$ can be obtained from the non-extremal
one given in \eqn{blackbrane-phi} by taking the extremal limit, i.e.,
$r_+ = r_-$. These branes preserve one half of the space-time supersymmetries
and the dilaton at a given $\rho$ is related to its asymptotic value via
the relation given in \eqn{blackbrane-phi}. The (arbitrary) fixed value at
the wall of the cavity is therefore related to the (arbitrary) asymptotic
value of the dilaton with the fixed cavity size. So in the present context,
one can re-phrase the attractor mechanism as: the value of the dilaton
at the horizon
is independent of the fixed value of the dilaton at the wall of the cavity. So if
the attractor mechanism is at work, the wall of the cavity cannot be too close
to the horizon since otherwise the approach will be inconsistent with this
mechanism or it will give rise to a jump of the value of the dilaton
in the bulk which is
impossible. Fortunately, the situation is not so restrictive as indicated above.
Once the attractor mechanism is at work, any given point outside the horizon
has actually an infinite physical distance away from the horizon (see,
for example, \cite{Kallosh:2006bt,Goldstein:2005hq,Garousi:2007zb}) and so the
condition $\rho_B > r_+ = r_-$ with fixed $\rho_B$, the radial location of
the wall of the cavity, will be enough.

\section{Stability analysis of the black $p$-branes}

For the purpose of the stability analysis we will define some new
parameters following
refs.\cite{Braden:1990hw,Lundgren:2006kt,Carlip:2003ne} and rewrite
the action, the potential and the inverse temperature in terms of
those parameters. Let us first define the reduced charge as,
\be\label{reducedcharge} Q_d^{\ast} = \left(\frac{\sqrt{2}\kappa
Q_d}{\Omega_{\tilde d +1} \tilde d}
e^{a\bar\phi/2}\right)^{\frac{1}{\tilde d}} \ee Then from
\eqn{rminus} we have \be\label{newrminus} \bar r_- =
\frac{(Q_d^{\ast})^2}{\bar r_+} \ee Next we define the following
parameters, \be\label{parameters} x = \left(\frac{\bar r_+}{\bar
\rho_B}\right)^{\tilde d}, \qquad \bar b = \frac{\b}{4\pi\bar
\rho_B}, \qquad q = \left(\frac{Q_d^{\ast}}{\bar
    \rho_B}\right)^{\tilde d}
\ee
where the dimensionless parameter $\bar b$ is fixed, but the other two
parameters $x$ and $q$ vary. Note that the parameter $\bar b$ is related to
the inverse of temperature of the environment, $q$ is related to the charge
and $x$ is related to the horizon size. In terms of these parameters we have
\bea\label{parameters1}
\Delta_+ &=& 1 - \frac{\bar r_+^{\tilde d}}{\bar \rho_B^{\tilde d}} = 1 - x\nn
\Delta_- &=& 1 - \frac{\bar r_-^{\tilde d}}{\bar \rho_B^{\tilde d}} =
1 - \frac{(Q_d^{\ast})^{2\tilde d}}{\bar r_+^{\tilde d} \bar \rho_B^{\tilde
    d}} = 1 - \frac{q^2}{x}\nn
1 - \frac{\bar r_-^{\tilde d}}{\bar r_+^{\tilde d}} &=& 1 -
\frac{q^2}{x^2} \eea Using \eqn{parameters1} the two equations at
the equilibrium (at the stationary point) giving the constant
temperature and the potential at the wall of the cavity given in
\eqn{didrplus} and \eqn{didq} can be written as \be\label{eos} \bar
b = b(x,q), \qquad \bar \varphi = \varphi(x,q) \ee where we have
defined $\bar \varphi = \sqrt{2}\kappa e^{-a\bar \phi/2} \bar\Phi$
and \be\label{bvarphi} b(x,q) = \frac{1}{\tilde d}
\frac{x^{\frac{1}{\tilde d}} (1-x)^{\frac{1}{2}}} {\left(1 -
\frac{q^2}{x^2}\right)^{\frac{1}{2}-\frac{1}{\tilde d}} \left(1 -
    \frac{q^2}{x}\right)^{\frac{1}{\tilde d}}}, \qquad
\varphi(x,q) = \frac{q}{x} \left(\frac{1-x}{1-
    \frac{q^2}{x}}\right)^{\frac{1}{2}}. \ee
The two equations in \eqn{eos} can also  be derived from the
following reduced action in a similar fashion as \eqn{didrplus} and
\eqn{didq} but now with respect to variables $x$ and $q$,
respectively, \bea\label{reducedaction} \tilde I_E (x, q) &\equiv&
\frac{2 \kappa^2 I_E}{4\pi \bar \rho_B^{\tilde d + 1} V_p
\Omega_{\tilde d + 1}} \nn &=& -  \bar b \left[(\tilde d + 2)
\left(\frac{1 - x}{1 - \frac{q^2}{x}}\right)^{\frac{1}{2}} + \tilde
d (1 - x)^{\frac{1}{2}} \left(1 - \frac{q^2}{x}\right)^{\frac{1}{2}}
- 2 (\tilde d + 1) + \tilde d q \bar \varphi\right]\nn & & \qquad -
x^{1 + \frac{1}{\tilde d}} \left(\frac{1 - \frac{q^2}{x^2}}{1 -
\frac{q^2}{x}}\right)^{\frac{1}{2} + \frac{1}{\tilde d}},\eea
where $I_E$ is as given in \eqn{ctocgaction2}.
These
two equations, for given $\bar b$ and $\bar \varphi$, determine $x$
and $q$ completely. However, in the presence of two variables, the
analysis of stability at the extremal points determined by these two
equations is a bit more complicated than that given in
\cite{Lu:2010xt}, in the canonical ensemble. There the relation at the
charge equilibrium for fixed charge has been employed to reduce the two
variables to only one\footnote{While the one-variable stability analysis
given in \cite{Lu:2010xt} appears natural in the canonical
ensemble, there exists also a more general two-variable analysis
which is more complete yet more complicated and gives the same
stability condition. For the present case, the two-variable analysis
seems more natural but as we will show in the text that when the second
equation in \eqn{eos} for the fixed potential is used to reduce the
two variables to only one, the same stability condition can be
obtained.}. While the stability analysis with two variables, which
we will perform in the Appendix, is more complete, there actually
exists an analogous simpler analysis as in the canonical case. But for
this we need to employ the potential relation at the equilibrium for fixed
potential, as given in the second equation of \eqn{eos}, to reduce the
two variables to one variable. In what follows, we will first
perform the simpler one-variable analysis and find the stability
condition. We will perform the more complete yet more complicated
two-variable analysis in the Appendix and show that these two
approaches give the same stability condition.

 Note that since $x/q = (\bar
r_+/Q_d^{\ast})^{\tilde d} = (\bar r_+/\bar r_-)^{\tilde d/2} \ge
1$, so, $x \ge q$ and since $x=(\bar r_+/\bar \rho_B)^{\tilde d} \le
1$, so, $x$ has the range $q \le x \le 1$. The condition \eqn{weakc}
may put a more strict lower bound for $x$ if $q$ is too small. The
end point $x = 1$ corresponds to taking $\bar \rho_B = \bar r_+$,
i.e., the cavity is placed on the horizon where the ensemble
temperature is infinity. In practice, we can keep an ensemble at any
given high temperature and the infinitely high temperature should be
understood as a limiting process. So we should take $x < 1$.
The other end point $x = q$ corresponds to the extremal case which
should be excluded from the consideration  given what has been
discussed in the Introduction. For this reason, we take $x > q$.
So, following \cite{Braden:1990hw}, we limit ourselves
to the physical region of $0\leq q < x <1$ in what follows,
keeping in mind the condition \eqn{weakc}\footnote{As will become
clear, the local stability condition given in \eqn{condition2b} does
satisfy the condition \eqn{weakc}.}. Also note from $\varphi$
expression in \eqn{bvarphi} that since $1-q^2/x = 1-(q^2/x^2)x>1-x$,
so we have $0\leq \varphi < 1$, with $\varphi =0$ corresponding to
the chargeless ($q=0$) case. As discussed in Introduction, we should
exclude the $\varphi = 1$ ($x = q$) since it corresponds to the
extremal case. $\varphi > 1$ cannot be in a black brane phase and it
actually corresponds to the `hot flat space' phase.

Let us now consider the one-variable case first, i.e., when the
second equation in \eqn{eos} $0\le \bar\varphi = \varphi (x, q) < 1$
is satisfied. We can now solve this equation to get, \be
\label{qoverx}\frac{q^2}{x^2} = \frac{\bar\varphi^2}{1 - (1 -
\bar\varphi^2) x}.\ee Now substituting \eqn{qoverx} in $b (x, q)$
equation in \eqn{bvarphi} we have for fixed potential $\bar\varphi$,
\be \label{onevb} b (x) \equiv b (x, q) = \frac{x^{\frac{1}{\tilde
d}}\left[1 -
    \left(1-\bar\varphi^2\right)x\right]^{\frac{1}{2}}}{\tilde d
  \left(1-\bar\varphi^2\right)^{\frac{1}{2}-\frac{1}{\tilde d}}},
\ee and in the reduced action in \eqn{reducedaction}, we have  \be
\label{fraction} \tilde I_E (x) \equiv \tilde I_E (x, q) = - 2 \bar
b \left(\tilde d + 1\right)\left[\left(1 - \left(1 -
\bar\varphi^2\right) x\right)^{1/2} - 1\right] - x^{1 + 1/\tilde d}
\left(1 - \bar \varphi^2\right)^{1/2 + 1/\tilde d}.\ee One can
easily check that \be \frac{d \tilde I_E (x)}{d x} = \frac{(1 +
\tilde d) (1 - \bar\varphi^2)}{\left[1 - (1 -
\bar\varphi^2)x\right]^{1/2}} \left[\bar b - b (x)\right],\ee where
$b (x)$ is defined in \eqn{onevb}. This vanishes at the stationary
point of the action, i.e., \be \label{presenteos} \bar b = b (\bar
x),\ee which is the first equation in \eqn{eos} in the present
consideration. The local stability at the stationary point is
determined by requiring it to be a local minimum of the action,
i.e.,  \be \left.\frac{d^2 I_E (x)}{d x^2}\right|_{x = \bar x} = -
\frac{(1 + \tilde d) (1 - \bar\varphi^2)}{\left[1 - (1 -
\bar\varphi^2)\bar x\right]^{1/2}} \frac{d b (\bar x) }{d \bar x}
> 0.\ee In other words, we need to have (since the other factors are
all positive) \be \frac{d b(\bar x)}{d\bar x} = \frac{{\bar
x}^{1/\tilde d - 1}\left[2 - (2 + \tilde d)\bar x (1 -
\bar\varphi^2)\right]}{2 \tilde d^2 (1 - \bar\varphi^2)^{1/2 -
1/\tilde d}\left[1 - (1 - \bar\varphi^2)\bar x\right]^{1/2}} < 0,
\ee which gives the local stability condition as \be
\label{localstability} 2 - (2 + \tilde d) \bar x (1 - \bar\varphi^2)
< 0.\ee In the above, $\bar x$ is a solution of the present equation
of state as given in \eqn{presenteos}.

As we will show in detail in the Appendix that the above local
stability condition \eqn{localstability}
continues to hold when we consider the more
general yet more complicated two-variable situation and this
justifies the simplified one-variable analysis of
stability performed above.

We therefore conclude that for the local stability of the system we
must have \eqn{localstability} to be satisfied. To understand the
meaning of the stability condition \eqn{localstability} we rewrite
it as \be\label{condition2a} \bar x \left(1 - \bar \varphi^2\right)
> \frac{2}{\tilde d +2} \ee which further implies,
\be\label{condition2b} \frac{2}{\tilde d +2} < \bar x < 1, \qquad
{\rm and} \qquad 0 < \bar \varphi < \sqrt{\frac{\tilde d }{\tilde d
+2}}. \ee In other words, for given $\bar \varphi$ satisfying the
above constraint, only $\bar x$ which is in the range specified in
\eqn{condition2b}, gives at least a locally stable system. Otherwise
the system is not stable. Note that the lower bound $2/(\tilde d +
2)$ for  $\bar x$ is of the order of unity for allowed $\tilde d \le
7$ and the condition \eqn{weakc} in the present context is $  l/\bar
\rho_B \ll 1$. So we can easily have $1 > \bar x > 2/(\tilde d + 2)
\gg l/\bar\rho_B$ to satisfy the condition \eqn{weakc}.

This concludes our discussion on the stability of asymptotically
flat black $p$-branes in the grand canonical ensemble. Using this
information we will construct the phase structure of the equilibrium
states of the black $p$-branes in this ensemble and compare with
that in the canonical ensemble.

\section{Phase structure of black $p$-branes}

We have seen in the previous section that the stability condition
for the black $p$-branes is given by \eqn{localstability}. When we
use the potential condition at the equilibrium, $\bar \varphi =
\varphi (x, q)$, to eliminate $q$ and obtain \eqn{onevb}, i.e, \be
\label{binxvarphi1} b (x)  = \frac{x^{\frac{1}{\tilde d}}\left[1 -
    \left(1-\bar\varphi^2\right)x\right]^{\frac{1}{2}}}{\tilde d
  \left(1-\bar\varphi^2\right)^{\frac{1}{2}-\frac{1}{\tilde d}}},
\ee we have the range of $x$ as $0<x<1$. This is different from the
fixed charge case of the canonical ensemble where the range is
$q<x<1$, since here we are considering the grand canonical ensemble
where the potential $\bar\varphi$ is fixed and not the charge $q$.
Even for the zero charge case of the canonical ensemble, where the
range of $x$ in the two cases are the same, the phase structure in
these two cases, as we will see, will be quite different due to the
different boundary data in the two ensembles. Note that as $x \to
0$, $b (x\to 0) \to b (0)=0$, exactly as in the canonical case, but
as $x \to 1$, $b (x \to 1) \to b (1)$, where it has the value,
\be\label{b1} b (1) = \frac{\bar \varphi}{\tilde d \left(1-\bar
\varphi^2\right)^{\frac{1}{2} -\frac{1}{\tilde d}}} > 0 \ee and this
is different from the fixed charge canonical case including the zero
charge case. The local temperature at this point has the value
\be\label{t1} T (1) = \frac{\tilde d \left(1-\bar
\varphi^2\right)^{\frac{1}{2} -\frac{1}{\tilde d}}}{4\pi \bar \rho_B
\bar \varphi}. \ee Also note from \eqn{binxvarphi1} that in the
range $0<x<1$, $b (x) > 0$ and so, if we plot $b (x) $ vs. $x$ curve
it will start from zero when $x \to 0$ and then it rises and ends at
$b (1)>0$ given above when $x \to 1$. In between there can be
extrema for $b (x) $ and in fact, it is not difficult to check from
\be \label{dbdx} \frac{d b (x)}{d x} = \frac{b (x) \left[2 - (\tilde
d + 2) x (1 - \bar\varphi^2)\right]}{2\tilde d\, x \left[1 - x (1 -
\bar\varphi^2)\right]},\ee that there can be only one extremum and
that is a maximum if it exists at all. This maximum is not always
guaranteed to exist unlike the chargeless case of canonical
ensemble. However, when it exists indeed, the characteristic
behavior of $b (x)$ vs. $x$ is given in Figure 1. The plot of $T
(x)$ vs. $x$ curve will have opposite behavior. It will start from
infinity when $x \to 0$ then the curve will drop and end at $T (1)$
as $x \to 1$. In between there can be an extremum and in this case
it is a minimum. Let us now determine the condition for the
existence of the maximum in $b (x)$ or minimum of $T (x)$. Equating
\eqn{dbdx} to zero gives us the maximum of $x$ as, \be\label{xmax}
x_{\rm max} = \frac{2}{(2+\tilde d)(1 - \bar \varphi^2)}. \ee Now
substituting this value in \eqn{binxvarphi1} we determine the
maximum value of $b (x)$ and from there the minimum value of $T (x)$
as, \bea\label{bmax} && b_{\rm max} = \left(\frac{2}{2+\tilde
d}\right)^{\frac{1}{\tilde d}} \left[\tilde d (\tilde d +2) \left(1
- \bar
    \varphi^2\right)\right]^{-\frac{1}{2}} \quad \Rightarrow \nn
&&T_{\rm min} = \left(\frac{2+\tilde d}{2}\right)^{\frac{1}{\tilde
d}} \frac{\left[\tilde d (\tilde d +2) \left(1 - \bar
    \varphi^2\right)\right]^{\frac{1}{2}}}{4\pi \bar \rho_B}.
\eea The condition for the existence of maximum is $x_{\rm max} <
1$, which gives $\bar \varphi < (\tilde d/(\tilde d +2))^{1/2}$,
i.e., the same constraint we obtained earlier in \eqn{condition2b},
which is expected since the non-existence of maximum implies no
local stability. From this we have the following constraint on
$b_{\rm max}$ or $T_{\rm min}$, \be\label{bmaxconstraint} b_{\rm
max} < \frac{1}{\sqrt{2\tilde d}}\left(\frac{2}{2+\tilde
    d}\right)^{\frac{1}{\tilde d}} \quad \Rightarrow
\quad T_{\rm min} > \frac{\sqrt{2\tilde d}}{4\pi \bar
\rho_B}\left(\frac{2+\tilde
      d}{2}\right)^{\frac{1}{\tilde d}}
\ee where we have used the constraint on $\bar \varphi$ given in
\eqn{condition2b}. So, when $b_{\rm max}$ ($T_{\rm min}$) given in
eq.\eqn{bmax} satisfy the constraint \eqn{bmaxconstraint} we will
have the maximum (minimum). From Figure 1, we can see that when
$x_{\rm max} < x < 1$, $d b /d x < 0$ and we have a locally stable
system and when $ 0< x < x_{\rm max}$, $d b/d x >0$ and we have an
unstable system consistent with what we discussed earlier.

\begin{figure}
 \psfrag{A} {$x$} \psfrag{B}{$1$} \psfrag{C}{$x_{\rm max}$}
 \psfrag{D}{$0$} \psfrag{E}{$b_{\rm max}$}
 \psfrag{F}{$b(x) $}\psfrag{G}{$\bar b$} \psfrag{H}{$x_1$}
 \psfrag{L}{$x_2$}
\begin{center}
  \includegraphics{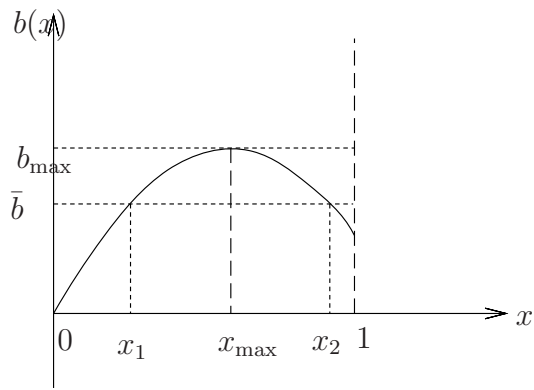}
  \end{center}
  \caption{The typical behavior of the (reduced) inverse temperature $b (x) $ vs. the (reduced) horizon radius $x$ at a fixed potential $\bar\varphi$.}
\end{figure}

For any given $\bar \varphi < 1$ and $\bar b$, satisfying $b
(1)<\bar b < b_{\rm max}$ with $b (1)$ as given in \eqn{b1}, one
expects two black brane solutions with radii $x_1$ and $x_2$, where
$x_1 < x_2$ from $\bar b = b (x)$ as shown in Figure 1 and only the
large one ($x_2$) will give a locally stable phase. The small one
($x_1$) will be unstable. Since here for each given $\bar \varphi$
and $\bar b$, we have only one locally stable phase, we don't have a
thermodynamical phase transition like in the canonical
ensemble\footnote{In the canonical ensemble there is a phase
transition between two stable black brane phases, one of which is
locally stable and the other is globally stable. Which one is
locally stable and which one is globally stable depend on the
temperature of the environment in contact with the system
\cite{Lu:2010xt}.}, even though $x_1=x_2=x_{\rm max}$ seems to
appear as a critical point. When $\bar \varphi \ge (\tilde d/(\tilde
d + 2))^{1/2}$ and/or $\bar b
> b_{\rm max}$ ($\bar T < T_{\rm min}$) no black brane phase is
possible in the grand canonical ensemble and the only possible
thermally stable phase here is the `hot flat space'. Also for $ 0 =
b (0)<\bar b < b (1)$, assuming $\bar \varphi< (\tilde d/(\tilde d +
2))^{1/2}$ from now on, no stable black brane phase is possible and
the `hot flat space' is again the thermally stable phase.  This
clearly shows that the thermodynamics and the phase structure are
quite different for the grand canonical ensemble than those in the
canonical ensemble \cite{Lu:2010xt} and this is expected since the
boundary data are different in the two cases.

To the leading order, the Euclidean action is directly related to
the grand potential or Gibbs free energy as $\beta \Omega = I_E$.
So, for a given temperature, smaller the value of $I_E$, the more
stable the system is. We will consider the reduced action $\tilde
I_E$. For a given $\bar b$ and $\bar \varphi$ satisfying their
respective constraints such that the system is locally stable, if
the reduced action is positive at the stationary point, it is
metastable since the reduced action has smaller value, i.e., zero,
for the `hot flat space' with the same boundary data and will make a
transition to this phase. In other words, if the system is initially
at the locally stable black brane phase, after some time, a phase
transition would occur so the black brane phase will become the `hot
flat space' phase with a different topology via a  phase transition.
So, for making sure that we have a stable black brane phase, we need
to look for condition such that the minimum of the reduced action is
actually a global minimum. This can be simply realized by conditions
such that the stationary action at the minimum is negative.

The reduced action at the stationary point can be expressed from the
action \eqn{fraction} as, \be\label{redaction} \tilde I_E =
\frac{\bar b}{y}(\tilde d +2) (1 - y)\left(y - \frac{\tilde
    d}{\tilde d +2}\right)
\ee where we have used $b (\bar x) =\bar b$ with $\bar x$ the
stationary point and $y = \sqrt{1 - \bar x (1 - \bar\varphi^2)}$.
Given the fact that $y<1$, so a negative reduced action requires,
\be\label{ycondition} y < \frac{\tilde d}{\tilde d +2} \ee which in
turn gives, \be\label{ycondition'} \bar x \left(1 - \bar
\varphi^2\right) > \frac{4(\tilde d +1)}{(\tilde d
  +2)^2},
\ee where we have used the expression of $y$ given above. Note that
this condition is consistent with the local stability condition
given in \eqn{condition2a} since $4(\tilde d +1)/(\tilde d +2)^2 >
2/(\tilde d +2)$. Also note that $4(\tilde d +1)/(\tilde d +2)^2 < 1
$ for all $\tilde d$ and therefore, there is no contradiction.
Actually, the constraint \eqn{ycondition'} gives the following new
constraints \be \label{globalc}  \frac{4 (\tilde d + 1)}{(\tilde d +
2)^2} < \bar x < 1,\qquad 0< \bar \varphi < \frac{\tilde d}{\tilde d
+ 2}.\ee So \eqn{ycondition'} or \eqn{globalc} is the condition for
global stability while \eqn{condition2a} or \eqn{condition2b} is
merely the local stability condition. If we denote \be\label{xg}
\bar x_g = \frac{4(\tilde d +1)}{(\tilde d +2)^2 \left(1-\bar
    \varphi^2\right)} \,( > x_{\rm max})
\ee then for $0 < \bar \varphi < \tilde d/(\tilde d +2)$ and $b (1)
< \bar b < b_g$ ($T_g < \bar T < T (1)$), where, \be\label{bg} b_g =
\frac{\left(4(\tilde d +1)\right)^{\frac{1}{\tilde d}}}{(\tilde d
+2)^{1
    + \frac{2}{\tilde d}} \sqrt{1- \bar \varphi^2}} < b_{\rm max}
\ee and \be\label{tg} T_g = \frac{(\tilde d +2)^{1
    + \frac{2}{\tilde d}} \sqrt{1- \bar \varphi^2}}
{4\pi \bar \rho_B \left(4(\tilde d +1)\right)^{\frac{1}{\tilde d}}}
> T_{\rm min}, \ee the reduced action has a global minimum at $\bar x_g < x_2 <
1$. In other words, this ensemble can give a sensible description of
black brane thermodynamics only when the above conditions are
satisfied. We would like to point out that when the maximum of $b
(x) $ exists (or minimum of $T (x)$ exists), the grand canonical
ensemble is in some sense similar to the chargeless case of
canonical ensemble except around $x=1$.

While the above gives a clear picture of phase structure for black
$p$-branes in the grand canonical ensemble, we can also consider the
dependence of the free energy on the temperature at fixed potential,
as in the usual approach,  to understand the same structure. For
this, we note the known results from the above: 1) When $\bar\varphi
> 1$, the only stable phase is the `hot flat space' and we have the
vanishing free energy ($\tilde I_E = 0$). 2) When $ (\tilde
d/(\tilde d + 2))^{1/2} \le \bar\varphi < 1 $, we have the (reduced)
on-shell free energy for the branes \be\label{rfreeenergy} \tilde
\Omega (\bar x) \equiv \frac{\tilde I_E (\bar x) }{\bar b} = (\tilde
d +2)\frac{1 - y}{y}\left(y - \frac{\tilde
    d}{\tilde d +2}\right) > 0,\ee where $y = \sqrt{1 - \bar x (1 - \bar\varphi^2)}$ and
     $(\tilde d/(\tilde d + 2))^{1/2}  < y < 1$. When this
     happens,  $b (x)$ does not have a maximum and there
     exists only one solution $\bar x$ from the equation $\bar b = b (\bar
     x)$ for $ 0 = b (0) < \bar b < b (1)$.
The reduced temperature defined as $t \equiv 1/\bar b =
     4\pi \bar \rho_B T$ is now in the range of $t (1) < t < \infty$
     with \be \label{rt1} t (1) = \frac{\tilde d \left(1-\bar
\varphi^2\right)^{\frac{1}{2} -\frac{1}{\tilde d}}}{ \bar \varphi},
\ee from \eqn{b1}, and the plot of the (reduced) free energy
     vs the (reduced) temperature gives only the unstable branch as
shown in Figure 2(a). For this case, we do not have stable brane
phase and the only stable phase is the `hot flat space'.
      Here we have $\tilde \Omega (t\to \infty) \to 0$ and
\be \label{freeenergy1} \tilde \Omega (t \to t (1)) \to (\tilde d +
2) \frac{1 - \bar\varphi}{\bar\varphi} \left(\bar\varphi -
\frac{\tilde d }{\tilde d + 2}\right),\ee which is positive for the
allowed range $(\tilde d/(\tilde d + 2))^{1/2} \le \bar \varphi <
1$. Let us examine the slope of $\tilde\Omega (t)$
with respect to $t$ in general. We have \bea \label{slope} \frac{d
\tilde \Omega (t)}{d t} &=& - \frac{1 - \bar \varphi^2}{2 y^3}
\left[ \bar x (1 - \bar\varphi^2) - \frac{2}{\tilde d + 2}\right]
\frac{d \bar x (t)}{d t}\nn &=& - \frac{\tilde d}{\tilde d + 2}
\frac{\bar x (t) (1 - \bar \varphi^2)}{t \,y (t)} < 0,\eea  where in
the second line we have used \be b (\bar x) t = \bar b\, t = 1,
\qquad  \frac{d b (\bar x)}{d \bar x} \frac{d\bar x (t)}{d t} = -
\frac {1}{t^2} < 0,\ee and \eqn{dbdx}. In other words, the slope is
always negative.  3) When $0 < \bar\varphi < (\tilde d /(\tilde d +
2))^{1/2}$, $ b (x)$ has a maximum and occurs at $x_{\rm max}$ given
in \eqn{xmax}. This gives a minimum (reduced) temperature as \be
t_{\rm min} = 1/b (x_{\rm max}) = \left(\frac{2+\tilde
d}{2}\right)^{\frac{1}{\tilde d}} \left[\tilde d (\tilde d +2)
\left(1 - \bar
    \varphi^2\right)\right]^{\frac{1}{2}}.\ee
\begin{figure}
 \psfrag{A}{$\tilde \Omega (t)$} \psfrag{B}{$t$} \psfrag{C}{$t (1)$}
 \psfrag{D}{$t_{\rm min}$} \psfrag{E}{$t (1)$}
 \psfrag{F}{$t_g$}\psfrag{G}{$(a)$} \psfrag{H}{$(b)$}
 \psfrag{K}{$(c)$}
 \begin{center}
  \includegraphics{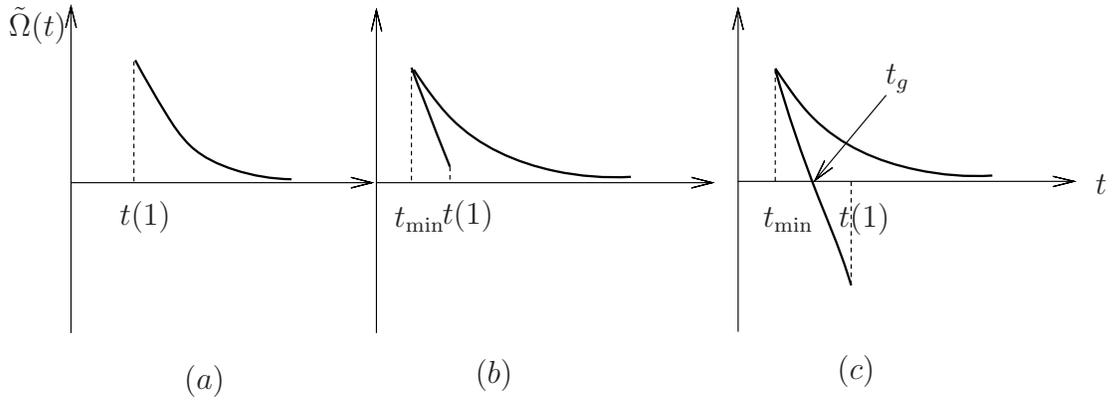}
  \end{center}
  \caption{The typical behavior of the (reduced) free energy $\tilde \Omega
(t)$ vs. the (reduced) temperature $t$ at a fixed potential
$\bar\varphi$.
  Plot (a) considers the case $ \sqrt{\tilde d/(\tilde d
    + 2)} \le \bar\varphi < 1$ which has only one unstable branch. Plot (b) considers
  the case $ \tilde d/(\tilde d + 2) \le \bar\varphi < \sqrt{\tilde d/(\tilde d
    + 2)}$ which has two branches, one unstable and the other only
locally stable branch.
  Plot (c) considers the case $0 < \bar\varphi < \tilde d/(\tilde d + 2)$
  which also has two branches, one unstable and the other (locally) stable
branch.
   The unstable branch has always positive free energy for
$t_{\rm min} \le t < \infty$ while the stable branch has negative
free energy in the range $t_g < t < t (1)$,
  where it becomes globally stable.}
\end{figure}There exist two solutions $ x_1 (t) $  and $x_2 (t)$ ($x_1 < x_2$)
from $ b (\bar x) = 1/t$
    for each given $t$ satisfying $ t (1) > t > t_{\rm
    min}$ as discussed earlier and shown in Figure 1. The small $x_1 (t)$
corresponds to the unstable brane and the
    large $x_2 (t)$ corresponds to the locally stable brane. This implies
that there
    always exist two branches in the plot of the (reduced) free energy
    vs the (reduced) temperature. The unstable branch is obtained by
    substituting the solution $x_1 (t)$ in the reduced free energy
    \eqn{rfreeenergy} with $t_{\min} <
    t < \infty$ while the locally stable branch is obtained
    similarly but now by substituting the solution $x_2 (t)$ instead
    with $t_{\rm min} < t < t (1)$, where $t(1)$ is again given by \eqn{rt1}
but $\bar\varphi$ has its present range
    as given above. The unstable branch always has positive free energy
which is
    larger than the free energy in the locally stable branch except
    at $t = t_{\rm min}$ where the two become equal and it is
    \be \label{mfe} \tilde \Omega (t = t_{\rm min}) = (\tilde d + 2)
\left(1 - \sqrt{\frac{\tilde d}{\tilde d + 2}}\right)^2 >0,\ee
    which is explicitly independent of $\bar\varphi$ in the present case.
The (reduced) free energy at the other end in this branch is
    again $\tilde \Omega (t \to \infty) \to 0$, also independent
of $\bar\varphi$. In this branch, we always have
    $x_1 (t) (1 - \bar\varphi^2) < 2/(\tilde d + 2)$ which can be
    obtained from $x_1 (t) < x_{\rm max}$ with $x_{\rm max}$ given in
    \eqn{xmax}. This branch gives no stable brane phase.
    For the locally stable branch, when $\tilde d/(\tilde d + 2) \le
    \bar\varphi
< (\tilde d/(\tilde d + 2))^{1/2}$,
    $\tilde \Omega (t) > 0$ for $t_{\rm min} < t < t (1)$ with the positive
$\tilde \Omega (t = t_{\rm min})$ given by \eqn{mfe} and the
     non-negative
    $\tilde\Omega (t \to t (1))$ still given by \eqn{freeenergy1} but now
for the present range of $\bar\varphi$ given above.
    The plot of the free energy vs. the temperature is given
    in Figure 2(b) for this subcase. This locally stable branch gives only
meta-stable brane phase and will eventually
    decay to the `hot flat space' after
    their formation.   If we want the existence of globally stable branes,
the free energy
    needs to be negative at certain temperature. This can be
    realized when $0 < \bar\varphi < \tilde d/(\tilde d + 2)$.
    With this, the free energy vanishes at $t = t_g$ and
    becomes negative when $t_g < t < t(1)$ with \be \label{rtg} t_g
    \equiv 1/b_g = \frac{(\tilde d +2)^{1
    + \frac{2}{\tilde d}} \sqrt{1- \bar \varphi^2}}
{ \left(4(\tilde d +1)\right)^{\frac{1}{\tilde d}}},\ee from
\eqn{bg}, therefore this branch becomes globally stable in this
range.  One can check that the (reduced) free energy $\tilde \Omega
(t \to t (1)) < 0$ as expected in the stable branch using its
expression given again by \eqn{freeenergy1}. Note that the value of
this limit increases as $\bar\varphi$ increases. One can also check
that $t_{\rm min} < t_g < t (1)$ in the present subcase. The plot of
the (reduced) free energy vs the (reduced) temperature for fixed
$\bar\varphi$ is now given in Figure 2(c). For each given $t$
satisfying $t_{\rm min} \le t < t (1)$ in the present two subcases,
the corresponding plot just reflects what has been described. In the
range $0 < \bar \varphi < \tilde d/(\tilde d + 2)$, $t_{\rm min},
t_g$ and $t (1)$ all decrease when $\bar\varphi$ increases which can
be seen from their respective expressions. Note that $\tilde \Omega
(t = t_{\rm min})$ in both branches and $\tilde \Omega (t \to
\infty) $ in the unstable branch are both independent of
$\bar\varphi$,  and $\tilde \Omega (t \to t (1))$ increases as
$\bar\varphi$ increases. As shown above, the slope of $\bar \Omega
(t)$ is negative in either branch. These features give the general
character of the free energy vs the temperature plot in both
branches for different fixed potentials. One may take Figure 2(b) as
the consequence of increasing $\bar\varphi$ from Figure 2(c).

\section{Conclusion}

To conclude, in this paper we have studied the equilibria and the
phase structure of the asymptotically flat dilatonic black
$p$-branes in a fixed cavity in arbitrary dimensions $D$ in a grand
canonical ensemble. For this, we have considered both the
temperature and the potential fixed at the wall of the cavity and
compared with our previous study of the same system in a canonical
ensemble [1] for which the charge enclosed in the cavity, rather
than the potential at the wall of the cavity, is fixed.  We
computed the Euclidean action corresponding to the dilatonic black
$p$-brane solution of $D$-dimensional supergravity with maximal
supersymmetry with proper consideration of the above fixed
quantities at the wall of cavity in this ensemble. To the leading
(or zeroth-loop)  order this action is related to the grand
potential or Gibbs free energy of the system and is an essential
entity for the stability analysis. This action has two variables,
namely, the horizon size $r_+$ (related to the variable $x$) and the
charge $Q_d$ (related to the variable $q$). At the stationary point
of this action, we derived the expected equations for which $x$ and
$q$ have to satisfy with both the temperature and the potential
fixed at the wall of cavity.  The second derivatives of the action
with respect to $x$ and $q$ at the stationary point can be used to
analyze the local thermal stability of the system in this ensemble.

Unlike in the canonical case,  the directly obtained action here is
a function of two variables and this makes the stability analysis a
bit involved. However,  there exists also an analog of one-variable
analysis as in the canonical case and we have shown in the text that
the condition so obtained for the stability of the equilibrium
states as given in \eqn{localstability} is the same as that of the
rather complicated two-variable case. For the two-variable analysis
given in the Appendix, we have made use of a trick to use the $b$,
$\varphi$ variables as defined in \eqn{bvarphi} instead of original
$x$, $q$ variables. This makes the analysis a bit simpler and
straightforward. The stability condition can also be expressed as
those given in \eqn{condition2b}. So, for a given potential
satisfying the constraint given in \eqn{condition2b}, only the
horizon size lying in the range also given in \eqn{condition2b} gave
locally stable system.

We then obtained the phase structure of the black $p$-branes in this
ensemble. We found that locally stable black brane phase exists only
when the minimum temperature of the system given in \eqn{bmax}
satisfies the constraint \eqn{bmaxconstraint}. In this case, below
the minimum temperature, there is no stable black brane phase and
only stable phase is the `hot flat space'. Above this temperature
and below the value $T (1)$ given in \eqn{t1}, there are two black
brane phases, the larger one is locally stable and the smaller one
is unstable. The locally stable black brane phase becomes globally
stable only above the temperature $T_g$ given in \eqn{tg} and below
$T(1)$. Below $T_g$ and above $T_{\rm min}$ (given in \eqn{bmax}),
the stable black brane is only locally stable and will eventually
make a transition to the `hot flat space'. Finally, above $T (1)$
there is no stable brane phase and the only stable phase is again
the `hot flat space' phase. We also commented on the similarity of
this phase structure with that of the zero charge canonical case
except at one end of the $x$ variable, namely, near $x = 1$. This
structure is reminiscent of the Hawking-Page phase transition of the
AdS, dS and flat black holes discussed in
\cite{Chamblin:1999tk,Carlip:2003ne,Lundgren:2006kt}. However,
unlike in the canonical case, here we found that at a given
temperature when the stable phase exists it does not come in pairs
-- only a single stable phase appears. So, there is no
thermodynamical phase transition between two stable black brane
phases analogous to the van der Waals-Maxwell liquid-gas phase
transition like that appeared in the canonical case.

\section*{Acknowledgements:}
We would like to thank the anonymous referee for various questions and
fruitful suggestions which help us to
improve the manuscript.
JXL acknowledges support by grants from the Chinese Academy of
Sciences, a grant from 973 Program with grant No: 2007CB815401 and a
grant from the NSF of China with Grant No : 10975129. Part of
the work was done when ZX was in the School of Physics and Astronomy
at the University of Southampton in UK.

\section*{Appendix: The two-variable stability analysis}
In this appendix we perform the more general yet more complicated
two-variable analysis of the stability of black $p$-branes
and show that the same local stability condition
derived from the simple one-variable consideration as given in
\eqn{localstability} continues to hold.

It is clear from \eqn{bvarphi} that both $b(x,q)$ and $\varphi(x,q)$
are smooth functions of $x$ and $q$ and so, in principle they can be
inverted to give $x$ and $q$ as functions of $b$ and $\varphi$.
Although we will not need their explicit form, however, we will need
the form of $b$ as a function of $x$ and $\varphi$. For that we will
eliminate $q$ from $b (x, q)$-equation given in \eqn{bvarphi} and
express it as a function of $x$ and $\varphi$ instead. From the
$\varphi$ equation in \eqn{bvarphi} we obtain $q^2/x^2$ as,
\be\label{qsquare} \frac{q^2}{x^2} = \frac{
\varphi^2}{1-\left(1-\varphi^2\right)x} \ee Now substituting
\eqn{qsquare} in $b$ equation in \eqn{bvarphi} we obtain,
\be\label{binxvarphi} b = \frac{x^{\frac{1}{\tilde d}}\left[1 -
    \left(1-\varphi^2\right)x\right]^{\frac{1}{2}}}{\tilde d
  \left(1-\varphi^2\right)^{\frac{1}{2}-\frac{1}{\tilde d}}}.
\ee The reduced action \eqn{reducedaction} can now be re-expressed
as, \be\label{twreducedaction} \tilde I_E (x, \varphi) = - \bar b
\left[(\tilde d + 2) y + \tilde d \left(1 - x (1 - \varphi\bar
\varphi)\right) y^{-1} - 2 (\tilde d + 1)\right] + b \tilde d (y^2 -
1) y^{-1}, \ee where we have used \eqn{qsquare} and
\eqn{binxvarphi}. In the above, we have also defined \be
\label{yfunction} y = \sqrt{1 - x \left(1 - \varphi^2\right)}, \ee
with $x(b,\varphi)$ determined by \eqn{binxvarphi}. We will use both
of \eqn{binxvarphi} and \eqn{twreducedaction} later.

We now expand the reduced action \eqn{reducedaction} at the
stationary point determined by the equation \eqn{eos} with $x=\bar
x$ and $q=\bar q$ to quadratic order as, \be\label{iexpansion}
\tilde I_E(x,q) = \tilde I_E(\bar x, \bar q) + \left.\tilde
  I_{ij}\right|_{z_i=\bar z_i, z_j=\bar z_j} (z_i-\bar z_i)(z_j-\bar z_j) +
\cdots \ee where $z_1 = x$, $z_2 = q$ with $i,\,j=1,\,2$ and we have
used the stationary conditions for the first order terms. In the
above \be\label{iij} \tilde I_{ij} \equiv \frac{\partial^2 \tilde
I_E(x,q)}{\partial z_i \partial
  z_j}
\ee The local stability of the system is determined by whether the
quadratic terms in the expansion are always positive definite. This
can be easily understood if we diagonalize the matrix $\tilde
I_{ij}$ and demand that each of the two eigenvalues $\lambda_1$ and
$\lambda_2$ is positive definite. Now since, \be\label{condition}
\lambda_1 \lambda_2 = {\rm det}\,\tilde I_{ij}, \qquad {\rm and}
\qquad \lambda_1 + \lambda_2 = \tilde I_{xx} + \tilde I_{qq} \ee
$\lambda_1,\,\lambda_2 > 0$ implies that both $\tilde I_{xx} +
\tilde I_{qq} > 0$ and ${\rm det}\,\tilde I_{ij} > 0$. Now we will
rewrite these two conditions such that our analysis becomes simpler
as\footnote{It is not difficult to see that the conditions given in
\eqn{condition1} and \eqn{condition1'} automatically implies ${\rm
det}\,\tilde I_{ij} > 0$ and $\tilde I_{xx}+\tilde I_{qq} >0$ . To
see this note that we first have to have $\tilde I_{qq}>0$ (if
$I_{qq} < 0$, the above two conditions can never be satisfied), then
\eqn{condition1'} implies ${\rm det}\, \tilde I_{ij}
>0$ and so, det $\tilde I_{ij} = \tilde I_{qq} \tilde I_{xx} - \tilde I_{qx}^2
= \tilde I_{qq}(\tilde I_{xx} - \tilde I_{qx}^2/\tilde I_{qq})>0$
$\Rightarrow$ $\tilde I_{xx} - \tilde I_{qx}^2/\tilde I_{qq} > 0$
$\Rightarrow$ $\tilde I_{xx}
> \tilde I_{qx}^2/\tilde I_{qq}>0$ and this in turn implies $\tilde I_{xx} +
\tilde I_{qq} > 0$.}, \bea \label{condition1} \tilde I_{qq} &>& 0,\\
\label{condition1'} \frac{\tilde I_{qq}}{{\rm det}\,\tilde I_{ij}}
&>& 0 \eea In order to understand the meaning of the condition
\eqn{condition1} we have to evaluate $\tilde I_{qq}$ and then set it
to greater than zero. From the form of $\tilde I_E$ in
\eqn{reducedaction} we obtain the first condition \eqn{condition1}
as, \bea\label{iqq}\tilde I_{qq}  &\equiv&  \left.\frac{\partial^2
\tilde I_E}{\partial q^2}\right|_{x=\bar x, q=\bar q}\nn &=&
\frac{\bar b \tilde d \bar \varphi}{\bar q \left(1 - \frac{\bar q
^2}{\bar x}\right)}\left[1 + \frac{(\tilde d +2)\bar
\varphi^2\left[1 - \frac{2}{\tilde d} + \frac{2}{\tilde d}\bar x
\left(1-\bar \varphi^2\right)\right]}{\tilde d \left(1-\bar
\varphi^2\right) (1-\bar x)} \right] > 0. \eea Once this is
satisfied the thermodynamic stability is completely determined by
the condition \eqn{condition1'}. A direct evaluation of $\tilde
I_{qq}/{\rm det}\,\tilde I_{ij}$ from the expression (in $(x,\,q)$
variables) of $\tilde I_E$ in \eqn{reducedaction} is complicated and
tedious and we will make use of a trick in evaluating it. This will
make the analysis of stability much more elegant and simpler. For
this,  we will instead evaluate the l.h.s. of \eqn{condition1'} by
first going to the $(b,\,\varphi)$ variable and evaluate $\tilde
I_{ab}$, where \be\label{iab} \tilde I_{ab} \equiv \frac{\partial^2
\tilde I_E(b,\varphi)}{\partial \xi_a
\partial \xi_b}
\ee with $a,\,b = 1,\,2$ and $\xi_1 = \varphi$, $\xi_2=b$ and then
relate this matrix to the original matrix $\tilde I_{ij}$ in
$(x,\,q)$ variables by chain rules as follows, \bea\label{iabiij}
\tilde I_{ab} &=& \left(\begin{array}{cc} \tilde
I_{\varphi \varphi} & \tilde I_{\varphi b}\\
\tilde I_{b \varphi}& \tilde I_{bb}\end{array}\right)\nn &=&
\left(\begin{array}{cc} \frac{\partial q}{\partial \varphi}&
\frac{\partial x}{\partial \varphi}\\
\frac{\partial q}{\partial b} & \frac{\partial x}{\partial
b}\end{array}\right) \tilde I_{ij} \left(\begin{array}{cc}
\frac{\partial q}{\partial \varphi}&
\frac{\partial x}{\partial \varphi}\\
\frac{\partial q}{\partial b} & \frac{\partial x}{\partial
b}\end{array}\right)^T, \eea where `$T$' denotes the transpose of a
matrix. After that we invert this matrix relation to obtain,
\be\label{iijiab} \tilde I_{ij}^{-1} = \left(\begin{array}{cc}
\frac{\partial q}{\partial \varphi}&
\frac{\partial x}{\partial \varphi}\\
\frac{\partial q}{\partial b} & \frac{\partial x}{\partial
b}\end{array}\right)^T \tilde I_{ab}^{-1} \left(\begin{array}{cc}
\frac{\partial q}{\partial \varphi}&
\frac{\partial x}{\partial \varphi}\\
\frac{\partial q}{\partial b} & \frac{\partial x}{\partial
b}\end{array}\right), \ee Again inverting \eqn{iijiab} we will
evaluate $\tilde I_{ij}/{\rm det}\,\tilde I_{ij}$ in terms of
$\tilde I_{ab}$ and other known functions given in \eqn{iijiab}. Now
to evaluate $\tilde I_{ab}$, we first calculate
\bea\label{delidelphi} \left(\frac{\partial \tilde I_E}{\partial
\varphi}\right)_b &=&
     \left(\frac{\partial \tilde I_E}{\partial x}\right)_q
     \left(\frac{\partial x}{\partial \varphi}\right)_b +
      \left(\frac{\partial \tilde I_E}{\partial q}\right)_x
\left(\frac{\partial q}{\partial \varphi}\right)_b \nn &=&
\left[\bar b - b (x, q)\right] f(x, q) \left(\frac{\partial
x}{\partial \varphi}\right)_b \nn &\,&- \left[\bar b \tilde d (\bar
\varphi - \varphi(x, q)) + \frac{(\tilde d + 2)\varphi (x, q)}{1 -
\frac{q^2}{x}} (\bar b - b (x, q))\right] \left(\frac{\partial
q}{\partial \varphi}\right)_b, \eea and \bea\label{delidelb}
\left(\frac{\partial \tilde I_E}{\partial b}\right)_\varphi &=&
     \left(\frac{\partial \tilde I_E}{\partial x}\right)_q
     \left(\frac{\partial x}{\partial b}\right)_\varphi +
      \left(\frac{\partial \tilde I_E}{\partial q}\right)_x
\left(\frac{\partial q}{\partial b}\right)_\varphi \nn &=&
\left[\bar b - b (x, q)\right] f(x, q) \left(\frac{\partial
x}{\partial b}\right)_\varphi \nn &\,&- \left[\bar b \tilde d (\bar
\varphi - \varphi(x, q)) + \frac{(\tilde d + 2)\varphi (x, q)}{1 -
\frac{q^2}{x}} (\bar b - b (x, q))\right] \left(\frac{\partial
q}{\partial b}\right)_\varphi, \eea where we have used the form of
$\tilde I_E$ given in \eqn{reducedaction} and the function $f(x,q)$
is defined as, \be\label{ffunction} f(x, q) = (1 - x)^{-\frac{1}{2}}
\left(1 - \frac{q^2}{x}\right)^{-\frac{1}{2}} \left[\tilde d + 2 -
\frac{\tilde d + 2}{2} \left(\frac{1 - \frac{q^2}{x^2}}{1 -
\frac{q^2}{x}}\right) + \frac{\tilde d}{2} \left(1 -
\frac{q^2}{x^2}\right)\right] > 0, \ee Thus we have from
eqs.\eqn{delidelphi} and \eqn{delidelb} \bea\label{doublederivs}
\frac{\partial^2 \tilde I_E}{\partial \varphi^2} = \bar b \tilde d
\frac{\partial q}{\partial\varphi}, \qquad \frac{\partial^2 \tilde
I_E}{\partial \varphi \partial b} = - f (\bar x, \bar
q)\frac{\partial x}{\partial \varphi}+ \frac{(\tilde d + 2)\bar
\varphi }{1 - \frac{\bar q^2}{\bar x}} \frac{\partial q}{\partial
\varphi},\nn \frac{\partial^2 \tilde I_E}{\partial b\partial
\varphi} = \bar b \tilde d \frac{\partial q}{\partial b}, \qquad
\frac{\partial^2 \tilde I_E}{\partial b^2} = - f (\bar x, \bar
q)\frac{\partial x}{\partial b} + \frac{(\tilde d + 2)\bar \varphi
}{1 - \frac{\bar q^2}{\bar x}}\frac{\partial q}{\partial b}, \eea
where the second derivatives are evaluated at $b=\bar b$ and
$\varphi = \bar \varphi$. So, from the above equations
\eqn{doublederivs} we obtain $\tilde I_{ab}$ as, \be
\label{matrixab} \tilde I_{ab} =\left(\begin{array}{cc} \bar b
\tilde d \frac{\partial q}{\partial \varphi} & \bar b \tilde d
\frac{\partial
q}{\partial b}\\
\bar b \tilde d \frac{\partial q}{\partial b} & - f (\bar x, \bar
q)\frac{\partial x}{\partial b} + \frac{(\tilde d + 2)\bar \varphi
}{1 - \frac{\bar q^2}{\bar x}}\frac{\partial q}{\partial
b}\end{array}\right), \ee where we have used \be \label{matrixabs}
\bar b \tilde d \frac{\partial q}{\partial b} = - f (\bar x, \bar
q)\frac{\partial x}{\partial \varphi}+ \frac{(\tilde d + 2)\bar
\varphi }{1 - \frac{\bar q^2}{\bar x}} \frac{\partial q}{\partial
\varphi}. \ee From the form of $\tilde I_{ab}$ given in
\eqn{matrixab} we can calculate its inverse as,
\be\label{inverseiab} \tilde I_{ab}^{-1} = \frac{1}{\det\tilde
I_{ab}} \left(\begin{array}{cc}- f (\bar x, \bar q)\frac{\partial
x}{\partial b} + \frac{(\tilde d + 2)\bar \varphi }{1 - \frac{\bar
q^2}{\bar x}}\frac{\partial q}{\partial b}  & - \bar b \tilde d
\frac{\partial
q}{\partial b}\\
- \bar b \tilde d \frac{\partial q}{\partial b} &\bar b \tilde d
\frac{\partial q}{\partial \varphi}
\end{array}\right),
\ee with det $\tilde I_{ab}$ having the from (as can be seen from
\eqn{matrixab}) \be\label{detiab} {\rm det}\, \tilde I_{ab} = f\bar
b \tilde d\left(\frac{\partial q}{\partial b}
  \frac{\partial x}{\partial \varphi} - \frac{\partial q}{\partial
    \varphi}\frac{\partial x}{\partial b}\right)
\ee where we have made use of \eqn{matrixabs}. Now substituting
\eqn{inverseiab} in \eqn{iijiab} we obtain, \be\label{inverseiij}
\tilde I^{-1}_{ij} = \left(\begin{array}{cc} \frac{1}{\bar b \tilde
d} \frac{\partial q}{\partial \varphi} & \frac{1}{\bar b \tilde d}
\frac{\partial
x}{\partial \varphi}\\
\frac{1}{\bar b \tilde d} \frac{\partial x}{\partial \varphi} &
\frac{(\tilde d + 2)\bar \varphi}{f\bar b \tilde d\left(1 -
\frac{\bar q^2}{\bar x}\right)} \frac{\partial x}{\partial \varphi}
- \frac{1}{f} \frac{\partial x}{\partial b}\end{array}\right), \ee
where we also made use of \eqn{matrixabs}. Inverting
\eqn{inverseiij} we get \be\label{finaliij} \tilde I_{ij} = \det
\tilde I_{ij} \left(\begin{array}{cc}\frac{(\tilde d + 2)\bar
\varphi}{f\bar b \tilde d\left(1 - \frac{\bar q^2}{\bar x}\right)}
\frac{\partial x}{\partial \varphi} - \frac{1}{f} \frac{\partial
x}{\partial b}  & -\frac{1}{\bar b \tilde d} \frac{\partial
x}{\partial \varphi}\\
- \frac{1}{\bar b \tilde d} \frac{\partial x}{\partial \varphi}
&\frac{1}{\bar b \tilde d} \frac{\partial q}{\partial \varphi}
\end{array}\right).
\ee From \eqn{inverseiij} we also obtain, det $\tilde I_{ij}^{-1}$ =
(det $\tilde I_{ij}$)$^{-1}$ = ${\rm det}\,\tilde I_{ab}/(f \bar b
\tilde d)^2$. Now from \eqn{finaliij} we finally get, \be
\label{gcstabilityone} \frac{\tilde I_{qq}}{\det \tilde I_{ij}} =
\frac{(\tilde d + 2)\bar \varphi}{f\bar b \tilde d\left(1 -
\frac{\bar q^2}{\bar x}\right)} \frac{\partial x}{\partial \varphi}
- \frac{1}{f} \frac{\partial x}{\partial b}, \ee We can compute
$\partial x/\partial b$ and $\partial x/\partial \varphi$ at the
stationary point from \eqn{binxvarphi} and obtain,
\be\label{bphisign} \frac{\partial x}{\partial b} = \frac{2 \tilde d
\bar x \left[1 - \bar x (1 - \bar \varphi^2)\right]}{\bar b \left[2
- (\tilde d + 2) \bar x (1 - \bar \varphi^2)\right]},\quad
\frac{\partial x}{\partial \varphi} = - \frac{2\tilde d \bar x \bar
\varphi \left[1 - \frac{2}{\tilde d} + \frac{2}{\tilde d}\bar x (1
-\bar \varphi^2)\right]}{(1 - \bar \varphi^2)\left[2 - (\tilde d +
2) \bar x (1 - \bar \varphi^2)\right]}, \ee Substituting
\eqn{bphisign} in \eqn{gcstabilityone} we have from the condition
\eqn{condition1'}, \be\label{newcondition} \frac{\tilde I_{qq}}{\det
\tilde I_{ij}} = - \frac{2\tilde d \bar x \left[1 - \bar x (1 - \bar
\varphi^2)\right]}{f \bar b\left[2 - (\tilde d + 2) \bar x (1 - \bar
\varphi^2)\right]}\left[1 + \frac{(\tilde d + 2)\bar
\varphi^2\left[1 - \frac{2}{\tilde d} + \frac{2}{\tilde d}\bar x (1
-\bar \varphi^2)\right]}{\tilde d (1 - \bar \varphi^2)(1 - \bar x)}
\right] > 0. \ee Thus for the stability of the system the above
condition has to be satisfied. Now since the second factor in the
square bracket is positive definite by the first condition written
in \eqn{iqq} it is clear that for \eqn{newcondition} to hold the
denominator of the first factor has to be negative definite. This
gives, as promised, precisely the same condition as given in
\eqn{localstability}.

Now it can be checked that once \eqn{localstability} is satisfied
the first condition written explicitly in \eqn{iqq} is automatically
satisfied. To see this we look at the numerator of the second term
in the square bracket in the expression of $\tilde I_{qq}$ given in
\eqn{iqq}, \bea\label{cond} 1-\frac{2}{\tilde d} + \frac{2}{\tilde
d}\bar x \left(1 - \bar
  \varphi^2\right)&=& \frac{1}{\tilde d}\left[\tilde d - 2\left(1-\bar
    x\left(1 - \bar \varphi^2\right)\right)\right]\nn
&=& \left[1 - \bar x \left(1 - \bar \varphi^2\right)\right] +
\frac{1}{\tilde d} \left[(\tilde d +2) \bar x \left(1 - \bar
\varphi^2\right) - 2\right] \eea and this is positive definite since
the first term is positive definite and the second term is positive
definite by \eqn{localstability}. Therefore, $\tilde I_{qq}$ is
positive definite. This shows that \eqn{localstability} is
equivalent to both the stability conditions given in
\eqn{condition1} and \eqn{condition1'}.

\end{document}